\documentstyle[11pt,newpasp,twoside,epsf]{article}
\def\edcomment#1{\iffalse\marginpar{\raggedright\sl#1\/}\else\relax\fi}
\reversemarginpar
\begin{document}
\title{Pulse Fluctuation Properties at 35~MHz}
\author{Ashish Asgekar$^{1,2}$ \& A. A. Deshpande$^{1}$}
\affil{$^{1}$Raman Research Institute, Sadashivnagar, Bangalore 560 080 INDIA.}
\affil{$^{2}$Joint Astronomy Programme, Indian Institute of Science, Bangalore~560~012 INDIA.}
\vspace{0.5cm}
\indent A few bright pulsars were observed at 35~MHz for $\ga{1000}$s using the
Gauribidanur Radio Telescope, and the data were analysed to study their
single-pulse fluctuation properties
(Asgekar \& Deshpande, 1999a; \& ref.s therein).\\
\indent The well-known drifter {\bf B0943+10} shows a well
resolved two-component profile at 35~MHz.
The longitude-resolved fluctuation spectrum (Fig 1)
shows a stable phase modulation
feature (aliased) at $0.459~c/P_{1}$, consistent with its drifting pattern
seen at higher radio-frequencies. Using the
``Cartographic Transform'' technique (Deshpande \& Rankin, 1999; hereafter DR),
we have mapped the pattern of its polar emission at 35 MHz
(Asgekar \& Deshpande, 1999b) as shown in fig 1.
Helped by the larger cone radius at lower frequencies,
the subbeams can be sampled in their full radial extent at 35 MHz.
These results combined with those  at higher frequencies (DR) suggest a
steadily rotating system consisting of 20
emission-columns in the polar region as that responsible for the
observed fluctuations over the entire range of observed frequencies.\\
\indent The fluctuation spectrum of
{\bf B0834+06} exhibits a feature at 0.461$c/P_{1}$ related to amplitude
modulation (fig 2) and has relatively low Q-value.
We estimate the circulation time associated with the underlying rotating
pattern of subbeams based on the fluctuation frequency (as well as phase)
and the viewing geometry for this pulsar.
This estimate depends crucially on the polarization PA-sweep rate and
we note that the polarization data presently available may be somewhat
unreliable. A preliminary polar-emission map made using our estimates
shows  distinct  subbeams delineating the emission cone (fig 2). More
importantly, the subbeam spacing is not as uniform as in the case of B0943+10,
consistent with the low Q of the features in the fluctuation spectra.\\
\indent  We have also examined the fluctuation spectra of
{\bf B1919+21} \& {\bf B0950+08} at 35 MHz.
In both cases, the fluctuation spectra agree well with
those seen at higher radio-frequencies. B0950+08 shows a featureless
spectrum, but with a bifurcated profile of modulated intensity.
A closer examination shows that more intense pulses from
B0950+08 seem to bifurcate preferentially.
\begin{figure}
\plottwo{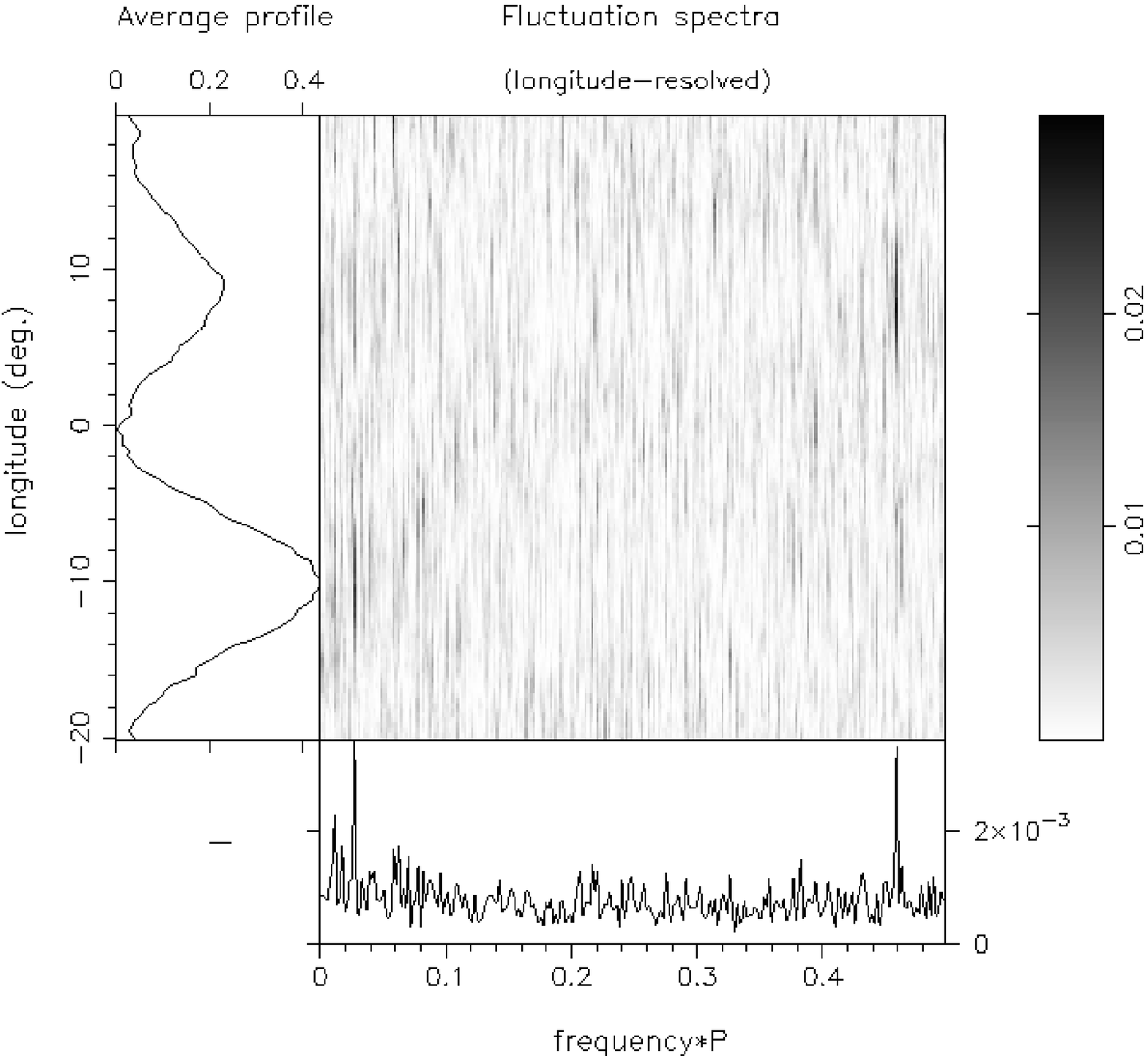}{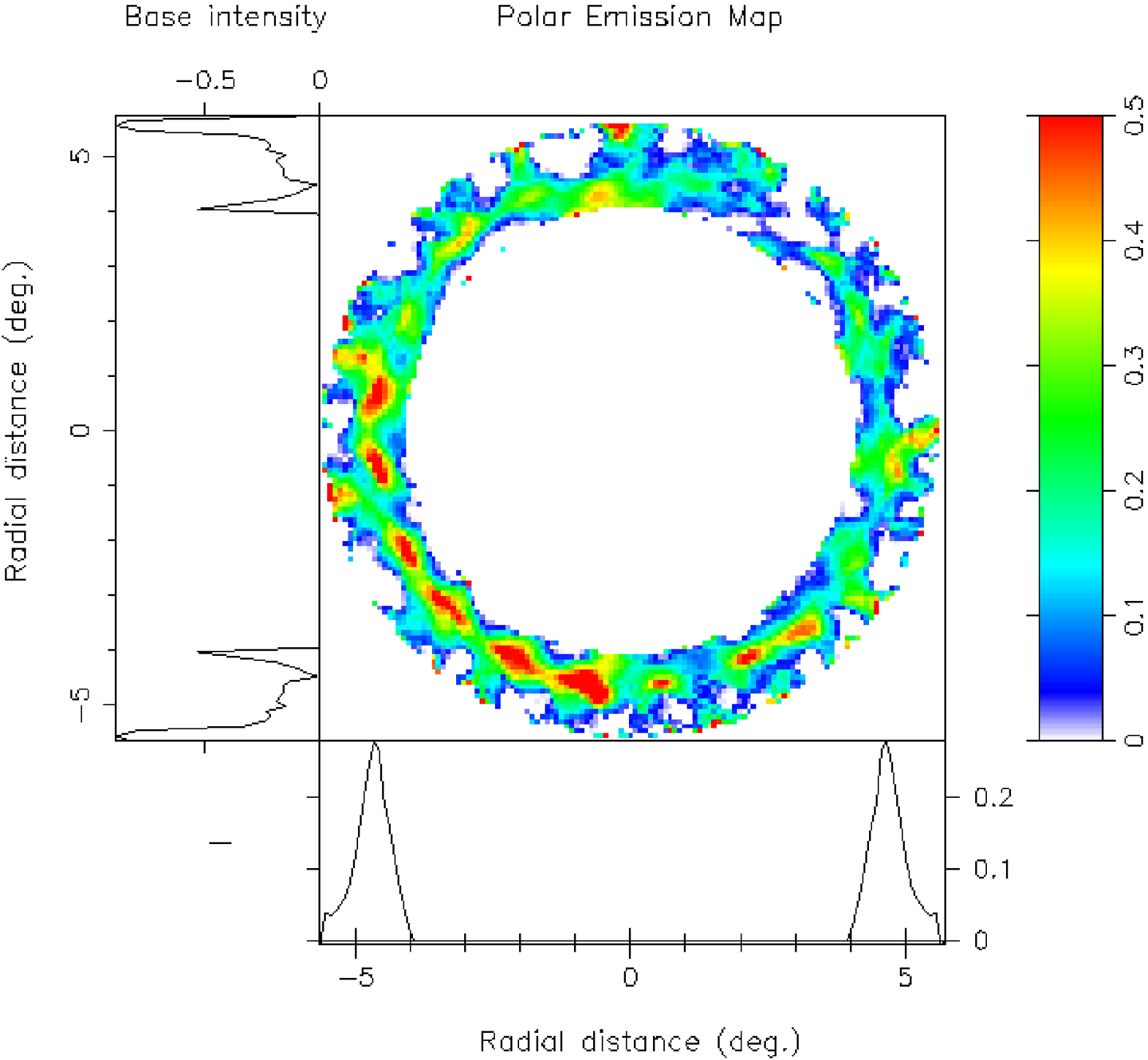}
\caption{Longitude-Resolved Fluctuation Spectrum (left) of B0943+10
\& the map of the polar emission pattern at 35 MHz.}
\end{figure}
\begin{figure}
\plottwo{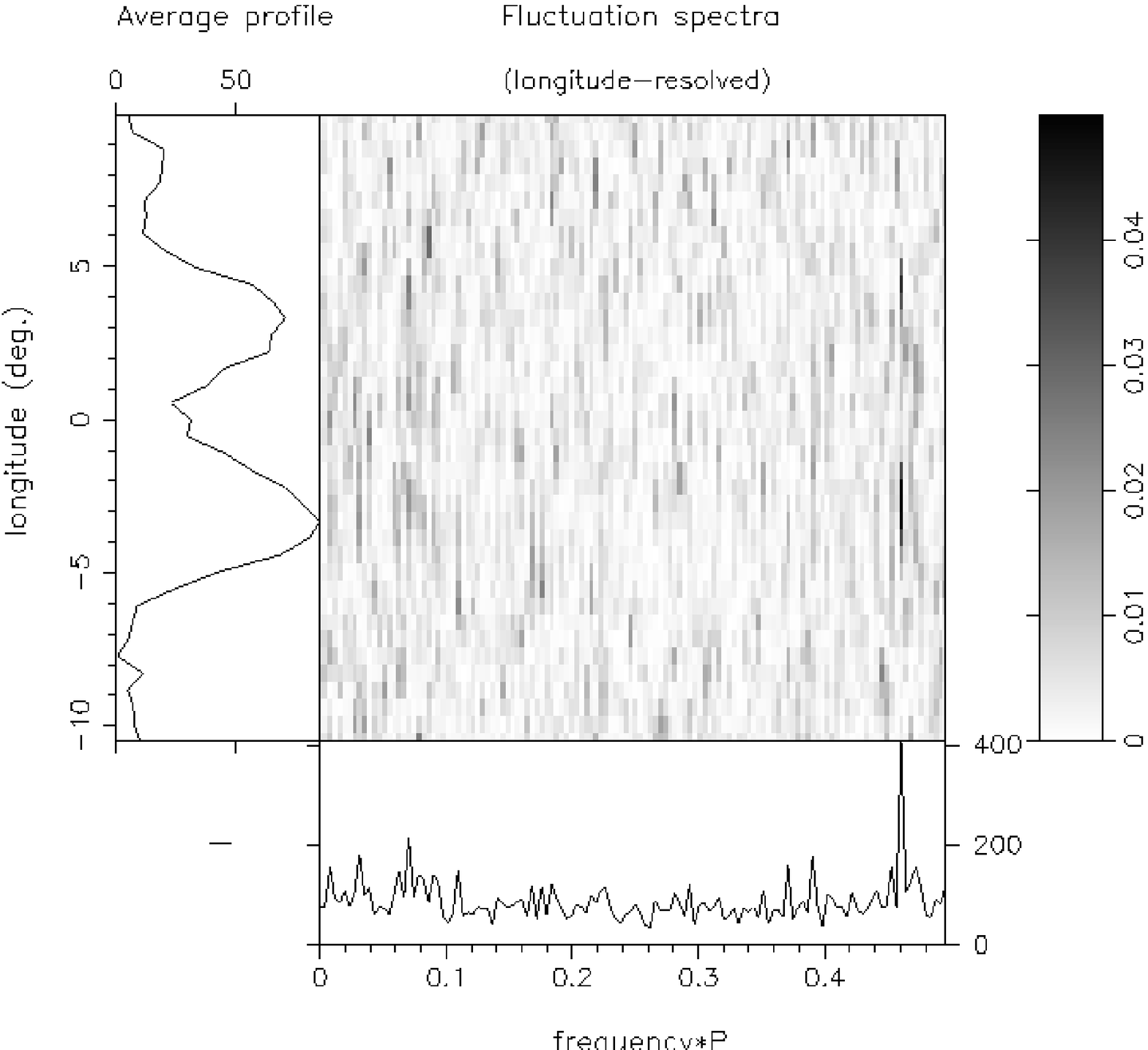}{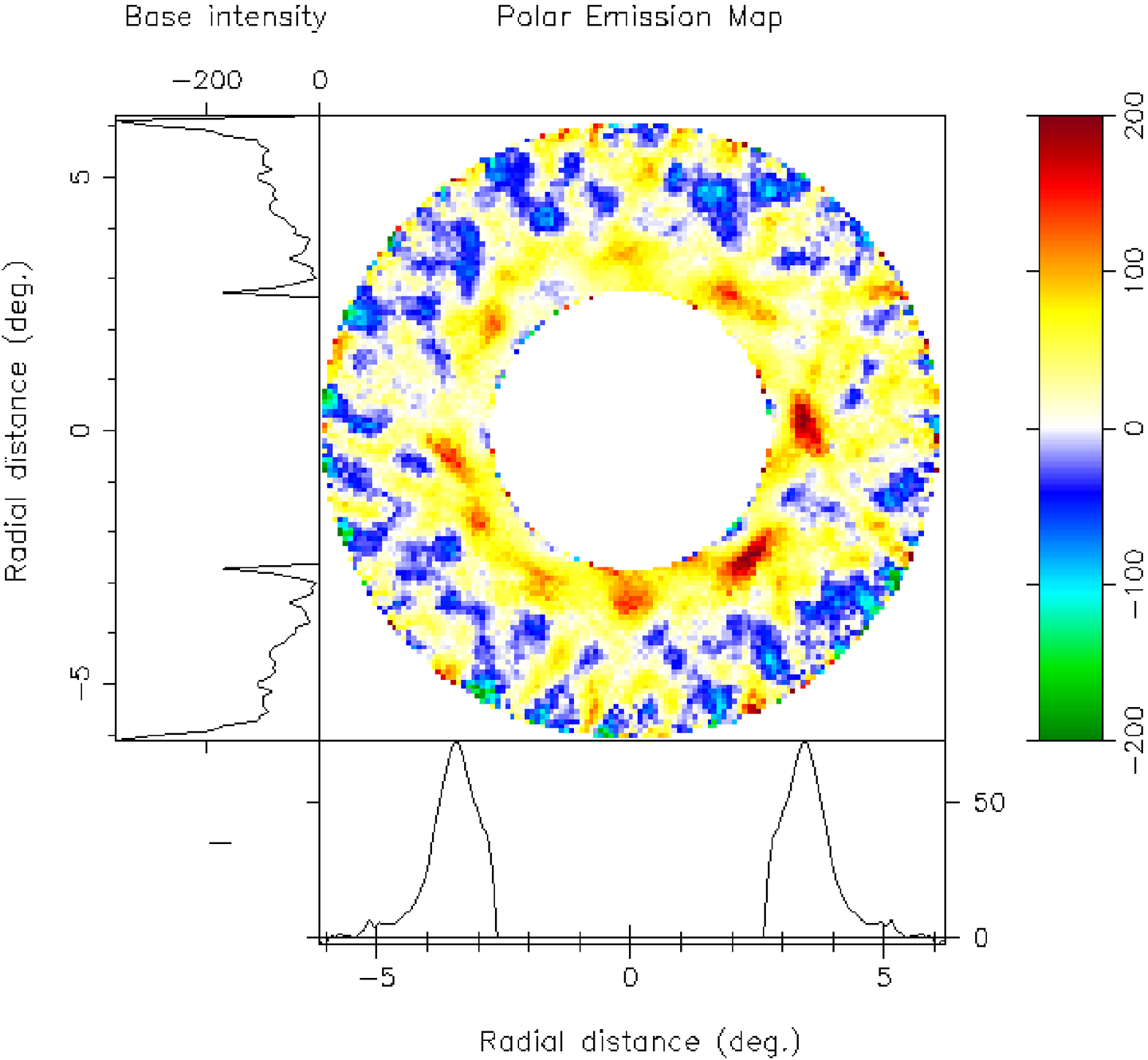}
\caption{Longitude-Resolved Fluctuation Spectrum~(left) \& polar emission
map of B0834+06 at 35 MHz. ($\alpha={30\deg}, \beta= {-3.0\deg}$)}
\end{figure}

\end{document}